\documentclass{ws-ijmpe}
\usepackage{epsfig,subfigure,latexsym}
\usepackage[super,compress]{cite}
\usepackage{amssymb,amsmath}
\newcommand {\bea}{\begin{eqnarray}}
\newcommand {\eea}{\end{eqnarray}}
\newcommand {\be}{\begin{equation}}
\newcommand {\ee}{\end{equation}}

\begin{document}

\def\({\left(}
\def\){\right)}
\def\[{\left[}
\def\]{\right]}

\def\Journal#1#2#3#4{{\it #1} {\bf #2}, (#4) #3}
\def\RPP{{Rep. Prog. Phys}}
\def\PRC{{Phys. Rev. C}}
\def\PRD{{Phys. Rev. D}}
\def\ZPA{{Z. Phys. A}}
\def\NPA{{Nucl. Phys. A}} 
\def\JPG{{J. Phys. G }}
\def\PRL{{Phys. Rev. Lett}}
\def\PR{{Phys. Rep.}}
\def\PLB{{Phys. Lett. B}}
\def\AP{{Ann. Phys (N.Y.)}}
\def\EPJA{{Eur. Phys. J. A}}
\def\NP{{Nucl. Phys}}  
\def\RMP{{Rev. Mod. Phys}}
\def\IJMPE{{Int. J. Mod. Phys. E}}
\def\AJ{{Astrophys. J}}
\def\AJL{{Astrophys. J. Lett}}
\def\AA{{Astron. Astrophys}}
\def\ARAA{{Annu. Rev. Astron. Astrophys}}
\def\MPLA{{Mod. Phys. Lett. A}}
\def\ARNPS{{Annu. Rev. Nuc. Part. Sci}}

\markboth{A. Sulaksono}
{Anisotropic pressure and hyperons in neutron stars}

\catchline{}{}{}{}{}

\title{ANISOTROPIC PRESSURE AND HYPERONS IN NEUTRON STARS}

\author{\footnotesize A. Sulaksono}

\address{\it Departemen Fisika, FMIPA, Universitas Indonesia,\\
Depok, 16424, Indonesia}

\maketitle   

\begin{history}
\received{(received date)}
\revised{(revised date)}
\end{history}

\begin{abstract}
We study the effects of anisotropic pressure on properties of the
neutron stars with hyperons inside its core within the framework
of extended relativistic mean field.  It is found  that the main
effects of anisotropic pressure on neutron star matter is to  increase
the  stiffness of the equation of state,   which compensates for
the softening  of the EOS due to the hyperons.  The  maximum mass and
redshift predictions of anisotropic neutron star with hyperonic
core  are quite compatible with the result of recent observational
constraints if we use the parameter of anisotropic pressure model $h \le
0.8$~\cite{HB2013} and $\Lambda \le -1.15$~\cite{DY2012}. The radius of
the corresponding neutron star at $M$=1.4 $M_\odot$ is more than 13 km,
while the effect of anisotropic pressure on the minimum mass of neutron
star is insignificant. Furthermore, due to the anisotropic pressure in
the  neutron star, the maximum mass limit of higher than 2.1 $M_\odot$
cannot rule out the presence of hyperons
in the neutron star  core.

\end{abstract}  \ccode{PACS numbers:97.60.Jd;14.20.Jd;26.60.Kp}

\section{Introduction} \label{sec_intro} The most accurate measurement
in identifying the masses of neutron star (NS) is the number of
pulsars in the bound binary systems (neutron-neutron and neutron-white
dwarf systems). Based on a recent analysis on mass distribution of
the number of pulsars with secure mass measurement, $M_G$ $\sim$
2.1 $M_\odot$ can be considered as an established value of lower
bound on maximum mass $(M_{max})$ for NS ~\cite{KKYT2013}. Therefore,
the existence of more massive NSs is, in principle, possible. The NS
maximum mass establishment comes from the result of two accurate NS mass
measurements. The mass 1.97 $\pm$ 0.04 $M_\odot$ of pulsar J1614-2230
is measured from the Shapiro delay~\cite {Demorest10} and the mass
2.01 $\pm$ 0.04 $M_\odot$~\cite{Antoniadis13} of pulsar J0348+0432 is
measured from the gravitational redshift optical lines of its  white
dwarf companion. In addition, there are evidences that some black
widow pulsars might have higher masses. For example, pulsar B1957+20
reportedly has a mass of $M_G$ = 2.4 $\pm$ 0.12 $M_\odot$~\cite{KBK2011},
and even gamma-ray black widow pulsar J1311-3430~\cite{RFSC2012} has
higher mass than  B1957+20 but with less accuracy. This NS maximum
mass limit poses a tight constraint on the equation of state (EOS) of
dense matter in the NS core. However, we point out that over the time,
with new planned observatories and technology advancement in astrometry,
it is not impossible to have accurately measured pulsar mass of higher
than 2.1 $M_\odot$ in the future~\cite{MCM2013,CHZF2013,LP2010}. 
 For the latest
review on neutron star masses and their implications  we refer the
reader to Refs. 8-11. It is also worthy to note that actually accurate
measurements of the NS radii would also strongly constrain the properties
of the matter in NS core. Unfortunately, the analysis methods used to
extract NS radii from observational data still have high uncertainty
and mostly they come from systematics~\cite{MCM2013}. Furthermore,
the limits of recent observational radii from different sources
or even from the same source are often in contradictory one to
another~\cite{Bog2013,Gui2013,LS2013,Leahy2011,Ozel2010,Stein2010,Stein2013,Sule2011}.

The observational constraint on the lower bound of the maximum NS mass of 2.1 $M_\odot$  can be readily fulfilled by most models if NS core contains only nucleons
and leptons. For examples IUFSU parameter set~\cite{FHPS2010}
yields $M_{\rm max}$= 1.94 $M_\odot$ and BSP parameter set~\cite{ASR2012}
yields  $M_{\rm max}$= 2.02 $M_\odot$.  On the other hand, the presence
of only nucleons and leptons in NS core is physically not too realistic.
In general the nuclear models that are compatible with the experimental
data on hyper-nuclei predict the existence of hyperons in matter at the
density of exceeding 2-3 times  nuclear saturation density ( $\rho_0=
0.16 \text{fm}^{-3}$) ~\cite{ZH2013}. Furthermore, the presence of
exotic particles such as hyperons in NS core has important impact on
NS cooling~\cite{Lackey2006}.  The hyperonization of matters tends to
soften the EOS of NS core as the  energetic nucleons are replaced by
slow moving hyperons. Consequently, the predicted maximum mass of NS
with hyperons in its core is always smaller than that of NS without
hyperons~\cite{Lackey2006,JSB_AG}. We note that the Brueckner-Hartree-Fock (BHF) model~\cite{BS2011,SR2011} yields
M$_{\rm max}$ $\sim$ 1.3-1.4   M$_\odot$ if hyperons are included
in the EOS of NS core.  It is reported by the authors of Ref. 27
that 3-body force also cannot help much to increase the predicted
NS maximum mass, but it is shown recently by the authors of Ref. 28
that 3-body forces can increase the maximum mass significantly. In
relativistic mean field (RMF) models, the situation is quite similar,
$M_{\rm max}$ $\sim$ 2.1 $ M_\odot$ can be reached only by adjusting
the model parameters in the hyperon sector or modifying nonlinear in
the strange sector  or introducing hypothetical weakly interacting
light boson (WILB)~\cite{BHZBM2012,SA2012,JLC12,WCS2012,SMK2011}.
The puzzle that whether or not the hyperons are present in the NS core has triggered the
theoreticians to revisit the NS models (see  Refs. 34-40 for details). The anticipation of
accurate measurement of NS with the mass of greater than  2.1 $M_\odot$
possible in the future, the parameters and model adjustments in the
framework of RMF models with acceptable nuclear EOS at low and moderate
densities may no longer be the best way to handle maximum mass.

 Usually one assumes that the pressure in the NS is isotropic.
But, there are
arguments (see Ref. 41 and the references therein) that the  matter
pressure of NS may be slightly different among different directions
(anisotropic). This effect can be caused by many interrelated factors
such as the presence of strong magnetic and electric fields, boson
condensations, different kinds of phase transition, the existence
of solid core or super fluidity, etc. This is also supported by
Herrera and Santos stating that formally, the mixture of two fluids
is mathematically equivalent to an anisotropic fluid (see Ref. 41
and the references therein). Furthermore, it has been shown that at
high density the nuclear matter pressure can also be   anisotropic
(see for example Refs. 42-43). Since the pioneer work of
Bowers and Liang~\cite{BL1974}, there have been many works devoted to
studies of anisotropic spherical symmetric configurations.
Recent studies on the properties of anisotropic star can
be found for examples in Refs.1-2 while studies for other anisotropic
configurations can be found in Refs. 45,47-49. These studies reveal that
anisotropy may have effects on maximum equilibrium mass and gravitational
redshift.

In this work, we argue that ``hyperonization puzzle''  can be
solved by considering that the pressure in NS matter may be
anisotropic. We study the NS mass, radius and the
gravitational redshift of anisotropic NS within the framework
of the extended version of relativistic mean field (ERMF)
model~\cite{ASR2012,Furnstahl96,Furnstahl97}.

The paper is organized as follows. Sec.~\ref{sec_EOS}, describes the
brief outline of NS EOS. Sec.~\ref{sec_AP}, discussion on anisotropic
pressure configuration. Section ~\ref{sec_RD} is devoted to discussion
of the results. Finally Sec.~\ref{conclu} is conclusion.

\section{Equation  of State}
\label{sec_EOS}
In general, NS has three parts with different compositions and density ranges i.e., outer crust, inner crust and the core. In this work, the crust EOS by Miyatsu $et ~al$.~\cite{MYN2013} is taken to describe NS crusts. The core is assumed to be composed of baryons  and leptons and the corresponding EOS is calculated by using the ERMF model.  

The ERMF model includes contribution from the standard RMF nonlinear self-interaction for $\sigma$ and $\omega$ mesons as well as additional cross interaction terms for $\sigma$, $\omega$ and $\rho$ mesons. The detail of this model and the EOS derivation based on this model are well documented~\cite{Furnstahl96,Furnstahl97}. In the ERMF model, baryons interact through exchanges of $\sigma$, $\omega$, $\rho$ and $\phi$ mesons, while the baryons involved are nucleons ($N$=$p$ and $n$) and hyperons ($H$=$\Lambda$, $\Sigma$ and $\Xi$). Thus the total Lagrangian density, including leptons ($l$=$e$ and $\mu$) for calculating the EOS of NS core can be written as ~\cite{SA2012}
\begin{equation}
{\mathcal{L}} = {\mathcal{L}}^{\rm free}_{B} + {\mathcal{L}}^{\rm free}_{M} + 
  {\mathcal{L}}^{\rm lin}_{BM} + {\mathcal{L}}^{\rm nonlin}+{\mathcal{L}}^{\rm free}_{l},  
\label{Lag}
\end{equation}
where the free baryons Lagrangian density is,
\begin{equation}
{\mathcal{L}}^{\rm free}_{B}=\sum_{B=N,\Lambda,\Sigma,\Xi}\overline{\Psi}_B[i\gamma^{\mu}\partial_{\mu}-M_B]\Psi_B,
\end{equation}
Here,  $\Psi_B$ is baryons field and the sum is taken from $N$, $\Lambda$, $\Sigma$, and $\Xi$ baryons. The Lagrangian density for the free mesons involved is, 
\begin{eqnarray}
{\mathcal{L}}^{\rm free}_{M}&=&\frac{1}{2}(\partial_{\mu}\sigma\partial^{\mu}\sigma-m_{\sigma}^2\sigma^2)+\frac{1}{2}(\partial_{\mu}\sigma^*\partial^{\mu}\sigma^*-m_{\sigma^*}^2\sigma^{*2})\nonumber\\ &-&\frac{1}{4}\omega_{\mu\nu}\omega^{\mu\nu}+\frac{1}{2}m_{\omega}^2\omega_{\mu}\omega^{\mu}-\frac{1}{4}\phi_{\mu\nu}\phi^{\mu\nu}+\frac{1}{2}m_{\phi}^2\phi_{\mu}\phi^{\mu}\nonumber\\ &-&\frac{1}{4}\mathbf{\rho}_{\mu\nu}\mathbf{\rho}^{\mu\nu}+\frac{1}{2}m_{\rho}^2\mathbf{\rho}_{\mu}\mathbf{\rho}^{\mu}.
\end{eqnarray}
The $\omega^{\mu\nu}$, $\phi^{\mu\nu}$ and $\mathbf{\rho}^{\mu\nu}$ are field tensors
corresponding to the $\omega$, $\phi$ and $\rho$ mesons field, and can be defined as
$\omega^{\mu\nu}=\partial^{\mu}\omega^{\nu}-\partial^{\nu}\omega^{\mu}$, $\phi^{\mu\nu}=\partial^{\mu}\phi^{\nu}-\partial^{\nu}\phi^{\mu}$
and $\mathbf{\rho}^{\mu\nu}=\partial^{\mu}\mathbf{\rho}^{\nu}-
\partial^{\nu}\mathbf{\rho}^{\mu}$. The Lagrangian ${\mathcal{L}}^{\rm lin}_{BM}$ describing interactions among baryons through mesons exchange is, 
\begin{eqnarray}                   
{\mathcal{L}}^{\rm lin}_{BM}&=& \sum_{B=N,\Lambda,\Sigma,\Xi}\overline{\Psi}_B[g_{\sigma B} \sigma + g_{\sigma^* B} \sigma^* -\gamma_\mu g_{\omega B} \omega^\mu\nonumber\\&-&\frac{1}{2}\gamma_\mu g_{\rho B}\mathbf{\tau_B}\cdot \mathbf{\rho} ^\mu -\gamma_\mu g_{\phi B}\phi ^\mu ]\Psi_B,
\label{eq:Llin}
\end{eqnarray}
where $\tau_B$ is the baryons isospin matrices. The Lagrangian describing
mesons self interactions for $\sigma$, $\omega$, and $\rho$ mesons can be
written as,
\begin{eqnarray}
 {\mathcal{L}}^{\rm nonlin} &=& - \frac{\kappa_3 g_{\sigma N} m_{\sigma}^2}{6 m_{ N} } \sigma^{3}
                   - \frac{\kappa_4 g_{\sigma N}^2 m_{\sigma}^2}{24 m_{ N}^2 } \sigma^{4}+\frac{\zeta_0 g_{\omega N}^2}{24} 
                   {(\omega_{\mu}  \omega^{\mu})}^2\nonumber\\
&+& \frac{\eta_1 g_{\sigma N} m_{\omega}^2}{2 m_{ N} } \sigma \omega_{\mu}  \omega^{\mu}+\frac{\eta_2 g_{\sigma N}^2 m_{\omega}^2}{4 m_{ N}^2 }\sigma^{2} \omega_{\mu}  \omega^{\mu} 
\nonumber\\&+&\frac{\eta_{\rho} g_{\sigma N} m_{\rho}^2}{2 m_{ B} } \sigma
\mathbf{\rho}_{\mu} \cdot \mathbf{\rho}^{\mu} +\frac{\eta_{1\rho}g_{\sigma
N}^2m_{\rho }^{2}}{4m_N^2} \sigma^2\mathbf{\rho}_{\mu} \cdot \mathbf{\rho}^{\mu} \nonumber\\ &+&\frac{\eta_{2 \rho} g_{\omega N}^2 m_{\rho}^2}{4 m_{ N}^2 } \omega_{\mu}  \omega^{\mu} \mathbf{\rho}_{\mu} \cdot \mathbf{\rho}^{\mu}.
\label{eq:Lnlin}
\end{eqnarray}
 While the free leptons Lagrangian density is, 
\begin{equation}
{\mathcal{L}}^{\rm free}_{l}=\sum_{l=e^-, \mu^-}\overline{\Psi}_l[i\gamma^{\mu}\partial_{\mu}-M_l]\Psi_l.
\end{equation}
here $\Psi_l$ is the leptons (electron and muon) field. The nucleons coupling constant and nonlinear parameters (BSP parameter set) are taken from Ref. 21. To determine the vector part of hyperons coupling constant $g_{\omega H}$ and $g_{\phi H}$, we consider conventional prescription based on SU(6) symmetry~\cite{JSB_AG} i.e., 

 \begin{eqnarray}
\frac{1}{3}g_{\omega N}&=&\frac{1}{2}g_{\omega
\Lambda}=\frac{1}{2}g_{\omega \Sigma}=g_{\omega \Xi},\nonumber\\
g_{\rho N}&=&\frac{1}{2}g_{\rho \Sigma}=g_{\rho \Xi},~ ~ ~ ~ ~
~ g_{\rho\Lambda}=0, \nonumber\\ 2 g_{\phi \Lambda}&=&2 g_{\phi
\Sigma}=g_{\phi \Xi}=\frac{2 \sqrt{2}}{3}g_{\omega N},~ ~ ~ ~ ~ ~
g_{\phi  N}=0.  
\label{eq:su6}
\end{eqnarray} 

For the given values of $g_{\omega H}$, the scalar hyperons coupling strengths  $g_{\sigma H}$ are usually obtained from the potential depth of hyperons in the symmetric nuclear matter evaluated at the saturation density $\rho_0$ as,
\begin{equation} 
U_{H}^{(N)}(\rho_0)
= -g_{\sigma H}\sigma(\rho_0)+g_{\omega H}\omega(\rho_0),
  \label{eq:uyn} 
\end{equation}
where the values of experimentally potential depth $U_H^{(N)}$ at  $\rho_0$ are~\cite{JSB_AG}
\begin{eqnarray}
 U_{\Lambda}^{(N)} &=& -28 {\rm ~MeV}, \quad U_{\Sigma}^{(N)} = +30 {\rm ~MeV}  \nonumber\\
&{\rm and}&  \quad U_{\Xi}^{(N)} = -18 { \rm ~MeV}.  
\label{eq:depth}
 \end{eqnarray}

The constituents composition in NS core is determined using standard conditions i.e., chemical potential balance, charge neutrality and baryon density conservation. The total energy density ($\epsilon$) of NS core matter can be determined from the zero component of energy-momentum tensor ($T^{00}$) that is obtained from Eq.~(\ref{Lag}). The radial pressure $p$ can be obtained from the thermodynamic relation as
\begin{equation}
p = \rho^2 \frac {d(\epsilon/\rho)}{d\rho}, 
\end{equation}
where $\rho$ is baryon density.

\section{Anisotropic Pressure in Spherical Symmetric Neutron Star}
\label{sec_AP}
As it is mentioned in introduction, the possible sources of local pressure anisotropy in spherically symmetric gravitating bodies that consisting not only low but also high densities matters are well known since long times ago~\cite{BL1974}.  Note that local pressure anisotropy here means that the radial pressure $p$ differs from the tangential pressure $q$. Many works after that up to now have been devoted to study this effect and to investigate the microscopic origin for each particular mechanism to generate pressure anisotropy in  spherically symmetric gravitating bodies founded in the literature. Furthermore, reviews about many of the possible causes for the appearance of this effect and their main consequences are also already exist~\cite{HS1997,Ivanov2010}. However, it will be quite informative for the reader if the microscopic basis of anisotropy of the fluid pressure in spherically symmetric NS that produced from variety physical processes is briefly discussed. 

\begin{enumerate}

\item Electric field\\
In general, if matter is composed of several kind of particles with different masses and opposite electric charges like happens in compact stars, the presence of sharp discontinuity between surface of the star and vacuum should lead to a charge separation and generation of an electric field~\cite{MEG2010}. From conservative point of view, because the star is macroscopic object, the global charge neutrality must be fulfilled. Thus the positive charge from the baryons or quarks in a compact star should be balanced by negative charge of electrons. However, since electrons are light and only electromagnetically interacting, they will penetrate through the boundary and generate  a local charge unbalanced around  the star surface. The conditions leading to the generation of electric field at star boundary have been studied by the authors of Ref. 53. It is also reported that strange stars may be expected to carry huge electric fields on their surfaces~\cite{Alcock1986,Kettner1995,Usov2004} thus it is quite wonder, if the electric field do not appear in NS surface. While it is known electrostatic interactions indeed are important for the description of neutron star crusts where atomic nuclei are embedded in dense electron gas~\cite{BBP1971,BMG2007}. We need also to note that there is also a unconventional neutron star model proposed by using less stringent condition i.e., they used a local charge neutrality condition instead the global one so that the electric field has been already explicitly taken into account in the corresponding model since the beginning~\cite{BPRRX2012,BBRR2014,RRWX2014}. Furthermore, it is also claimed in Refs. 62-64 that for compact stars where the density is high and the relativistic effect is crucial, in principle, in the allowed net charge of a compact star; the star can take some more charge to be in equilibrium. We need also to note that in NS matter, Debye screening may be generated. It will ensure that electric fields are confined in microscopic length scale of 10-1000 fm. If we assume that the electric field is nonzero then the matter stress-energy tensor in the right hand side of Einstein equation  with certain energy density and pressure will add with the terms from electromagnetic field, where for the case static and spherical symmetric stars, only the components of the  Maxwell field $F^{01}$ and $F^{10}$ are survived. In this specific case the total radial and tangential pressures of the stars are not the same anymore i.e., ~\cite{Bekenstein1971,Ray2003,Arbanil2013} 
\be
q = p + \frac{1}{4 \pi} \frac{Q^2}{r^4}, 
\ee
where the total charge $Q$ that produce electric field can be obtained from following relation
\be
\frac{dQ}{dr} = 4 \pi \rho_c {[1-\frac{2 G M}{r}+\frac{G Q^2}{r^2}]}^{-1/2} r^2.
\ee
Here $\rho_c$ is charge density while $M$ is total mass of the star. However, the actual form of  $\rho_c$ profile in each compact star case is indeed not certainly known. People usually use the parametrized  form of $\rho_c$. For example, in the case electric field effect on strange star studied in Ref. 65, the authors used 
\be
\rho_c \equiv \frac{K}{4 \pi r^2}[ \delta(r-R^+)-\delta(r-R^-)],
\ee
to describe the charge distribution in that star. Here $K$ is  a parameter which identified the strength of the charge. However, in general the effect of electric field on standard picture of spherically symmetric NS properties are not too significant. For example, it is shown by the authors of Ref. 53 that under Newtonian gravitation approach, that the electrostatic and gravitational contributions become equal at  minimum particle density $N_{min}$ $\approx$ 6.26 $10^{36}$. This value corresponds to the mass $M_{min}$ about $10^{13}$ gr, i.e. almost 20 order of magnitude smaller than the maximum mass of NS. ( see the detail in Ref.53 and references therein). Note that the maximum allowed charge at the surface of NS with $M \approx M_{\odot}$ and radius R around 10 km, is $Q \lesssim $  $10^{20}$ C~\cite{Ray2003}. With this charge value, we can estimate that the expected observed neutron star anisotropy due to electric field is $q-p$ = $\frac{1}{4 \pi} \frac {Q^2}{R^4}$ $ \lesssim $ 5  $10^{-4}$ $\rm MeV fm^{-3}$. It is much less compared to the center pressure of this corresponding NS, i.e., $p_c \sim $ 20  $\rm MeV fm^{-3}$.
\item Magnetic field\\
It is known from observations that pulsars have the typical surface magnetic  field strength  around $10^{12}$-$10^{13}$ G~\cite{Taylor1993} while the magnetars have surface magnetic field strength in the range of $10^{14}$-$10^{15}$ G~\cite{TD1992,TD1996}. As also discussed in Ref. 67, the central magnetic field  strength of the magnetars might be  as high as $10^{18}$-$10^{19}$ G. However, the origin of strong magnetic fields in compact stars is still not too clearly known up to now. The accepted large magnetic field generation mechanism in magnetars is based on  amplification of a seed magnetic field owing to the rapidly rotating plasma of a protoneutron star. Nevertheless, this mechanism can not substantiate all of the features of the supernova remnants surrounding these objects (see Ref. 69 and references therein). Whatever the way to generate magnetic field in  spherically symmetric compact stars, the presence of large magnetic field  leads to the generation of pressure anisotropy, where if we neglected the small contribution from matter magnetization and assuming that the magnetic field in the radial direction, the total tangential pressure can be written as (see for example Ref. 70)
\be
q = p + \frac{B^2}{4 \pi}, 
\ee
where $B$ is the magnetic field in the corresponding star.  Usually the magnetic profile of  spherically symmetric NS is assumed to be density dependent and is parametrized as (see Ref. 70 and references therein)
\be
B(\rho) \equiv B_s + B_0[1-e^{-\alpha {(\frac{\rho}{\rho_0})}^\gamma}],
\ee
where $\alpha$, $\gamma$ are parameters, $\rho_0$ is saturation density, $B_s$ and   $B_0$ are surface and center magnetic fields of the star. However, we need to note, in previous studied that it is shown that the contribution of magnetic field in EOS of neutron star matter through magnetization is not too significant (see Ref. 70 and references therein). If we take $B \lesssim B_0$ $\approx $ $10^{18}$ G, the expected observed neutron star anisotropy due to magnetic field can be estimated as $q-p$ = $\frac {B^2}{4 \pi} \lesssim $ 2  $10^{2}$ $\rm MeV fm^{-3}$.  We need to note that the strong magnetic field can affect also the surface electric field of strange stars~\cite{ZW2006}. The presence of magnetic field in matter might be also lead to the polarization of matter. It was shown~\cite{FMS2002} that the spin polarization induces  a deformation of the Fermi spheres of nucleons with spins parallel  and opposite to the polarization axes. This feature can be related to the structure of the one-pion exchange contribution to a realistic nucleon-nucleon interaction. The deformation is identified by angle dependent of Fermi momentum. This deformation will generate also additional pressure anisotropy ( see detail discussion in Ref. 73). 
\item Beyond one fluid description\\
In conventional view, it is assumed that NS matter is an ideal one fluid composed by several kind of particles with different masses and opposite electric charges. However, if we consider  NS matter as many fluids then the pressure anisotropy may appear in non trivial way. As an illustration, let see the simplest case. If we assume that the NS matter composed by 2 ideal fluids with different four velocity ($u_{\mu}$ and $w_{\mu}$), fluid 1 composed by neutral particles and fluid 2 composed by charged particles~\cite{HS1997,TL2011,LA1986}. The total stress-energy tensor for this system becomes
\bea
T_{\mu \nu}&=&(\epsilon_1+ p_1) u_{\mu} u_{\nu}+p_1  g_{\mu \nu}\nonumber\\
           &=&(\epsilon_2+ p_2) w_{\mu} w_{\nu}+p_2  g_{\mu \nu}.
\label{eqTmn}
\eea
Note, here we used metric sign (-,+,+,+) instead (+,-,-,-) which is used in Refs. 41,47,75. Therefore, the sign in some terms  here is different to that presented in Refs. 41,47,74-75. Eq.(\ref{eqTmn}) can be significantly simplified by casting it into standard form of anisotropic fluids. It can be done by using following transformation~\cite{HS1997,TL2011,Bayin1982,LA1986}
\bea
u^{\mu *}&=&u^{\mu} cos \alpha+ {\[\frac{\epsilon_2+ p_2}{\epsilon_1+ p_1} \]}^{1/2}w^{\mu} sin \alpha\nonumber\\
w^{\mu *}&=&w^{\mu} cos \alpha- {\[\frac{\epsilon_1+ p_1}{\epsilon_2+ p_2} \]}^{1/2}u^{\mu} sin\alpha,
\label{Rota}
\eea
where this transformation satisfies $T^{\mu \nu} (u,w)$= $T^{\mu \nu} (u^*,w^*)$. Explicitly it can be done by choosing $u^{\mu *}$ and $w^{\mu *}$ such that one is time-like and the other is space like so that $u^{\mu *}w_{\mu *}$=0, we can obtain the rotation angle in Eq. (\ref{Rota}) as
\be
tan 2 \alpha= 2 \frac{{(\epsilon_2+ p_2)(\epsilon_1+ p_1)}^{1/2}}{(\epsilon_2+ p_2)-(\epsilon_1+ p_1)} u^{\mu }w_{\mu }.
\ee
Followed by defining the quantities
\bea
v^{\mu }&=&\frac{u^{\mu *}}{{(-u^{\nu *}u_{\nu *})}^{1/2}} ~~~~~~~~~~~~~
\kappa^{\mu }=\frac{w^{\mu *}}{{(w^{\nu *}w_{\nu *})}^{1/2}}\nonumber\\
\epsilon&=&T^{\mu \nu}v_{\mu }v_{\nu } ~~~~~~~~~~~~~
\Psi=T^{\mu \nu}\kappa_{\mu }\kappa_{\nu }\nonumber\\
\Pi&=&p_1+p_2,
\eea
respectively, then the $T^{\mu \nu}$ of two perfect fluids in Eq.(\ref{eqTmn}) can be written as $T^{\mu \nu}$ of one fluid with anisotropy pressure as ~\cite{HS1997,TL2011,Bayin1982,LA1986}
\be
T_{\mu \nu}=\epsilon v_{\mu} v_{\nu}+\Psi  \kappa_{\mu} \kappa_{\nu} +\Pi [g_{\mu \nu}+u_{\mu} u_{\nu}- k_{\mu} k_{\nu}].
\ee
Note, To calculate this effect explicitly, beside the EOS of the matter, we need to know the magnitude of the scalar product of four velocity $u_{\mu}$ and $w_{\mu}$. In principle, both velocities should be determined from other physics information where in some cases, it is unknown. For example, in Ref. 75 in the case of bosonic dark matter model, due to lack of such information, the authors used $u_{\mu} w^{\mu}$ as  a parameter and studied the effect of the magnitude of this variable to the properties of bosonic dark matter. For providing rough estimation of the anisotropy effect from this mechanism, we may assume that only neutrons and protons in dominate the contribution in energy density $\epsilon$ and pressure $\Pi$ of NS. Now we denote the  $\epsilon_1$ and pressure $p_1$ as the energy density and pressure for proton fluid and $\epsilon_2$ and pressure $p_2$ as the energy density and pressure for neutron fluid and $u^{\mu} w_{\mu } \approx 1 +\frac{b}{2}$, where the $b$ a number to identify the four-velocity difference between two fluids.  By using Eqs. (6-7) in Ref. 75, it can be obtained that
\be
\Psi-\Pi\approx \frac{b}{2}\frac{(\epsilon_1+p_1)(\epsilon_1+p_1)}{(\epsilon+ \Pi)},
\ee
if we approximate that $(\epsilon_1+p_1) \approx Y_p(\epsilon+ \Pi)$ and $(\epsilon_2+p_2) \approx (\epsilon+ \Pi)$, where $Y_p$ is proton fraction, then $\Psi-\Pi\approx Y_p \frac{b}{2}(\epsilon+\Pi)$. It is known that the actual values of  $Y_p$, $\epsilon$ and $\Pi$ depend on the EOS  model used. Parameter $b$ also controls the significance of the effect. If the difference in four-velocity between two fluid is large the effect becomes larger and if it is small the effect is small. To estimate the number, lets for example say that the center density and its corresponding energy density for  $M \approx M_{\odot}$ i.e.,   $\Pi_c \approx$ 20 MeV with $\epsilon_c \approx$ 200 MeV and $Y_p \sim$ 0.1 as well as taking $b \sim$ 0.02 than it leads to 10 $\%$ effect. Thus we may expect also a quite substantial effect may come from this mechanism.

\item Other sources of anisotropy\\
Anisotropy in NS fluid pressure  might be also yielded by the existence of a solid core or by the presence of superfluid and by pion condensation, as well as by different kind of phase transition (see Ref. 2 and references therein for more detail). Another source of pressure anisotropy may come also from the matter viscosity of NS (see Ref. 1 and references therein for more detail).
\end{enumerate} 

To this end, we need to emphasize here that the effect of local anisotropy in  NS have been studied~\cite{DY2012,HH75,HS76,MH22003}. However, the discussion of NS local anisotropy by using more realistic and up to date NS matter EOS as well as by connecting this matter to the context of ``hyperonization puzzle'' is not yet done before and this becomes the focus of this work.

To accommodate the anisotropic fluid assumption, we start by taking the stress-energy tensor as~\cite{HB2013,DY2012}:
\begin{eqnarray}
T_{\mu \nu}=\epsilon u_{\mu} u_{\nu}+p  k_{\mu} k_{\nu} + q [g_{\mu \nu}+u_{\mu} u_{\nu}- k_{\mu} k_{\nu}].
\label{ETEN}
\end{eqnarray}
Here $g_{\mu \nu}$ is the space-time metric,  $u_{\mu}$ is the fluid 4-velocity, $\epsilon$ is the total energy density,  $k_{\mu}$ is the unit radial vector where $u^{\mu}k_{\mu}$=0. At the center of symmetry, the anisotropic pressure must vanish since here $k_{\mu}$ is no longer defined. Note that $g_{\mu \nu}+u_{\mu} u_{\nu}- k_{\mu} k_{\nu}$ is the projection tensor onto the 2-surface orthogonal to  $k_{\mu}$  and  $u_{\mu}$. Here, we use standard metric for spherically symmetric space time i.e.,
\be
ds^2=-e^{2 \nu} dt^2 +e^{2 \lambda} dr+ r^2 (d \theta^2 + sin^2 \theta ~d\phi^2).
\ee

By inserting Eq.~(\ref{ETEN}) into the Einstein field equations 
\begin{equation}
R_{\mu \nu}-\frac{1}{2} g_{\mu \nu}= 8 \pi G T_{\mu \nu},
\label{EQ}
\end{equation}
and followed by manipulating four equations from non-zero diagonal components of Eq.~(\ref{EQ}), we can obtain the Tolman-Oppenheimer-Volkoff (TOV) equations for anisotropic star as follows
\begin{equation}
\frac{dp}{dr} = -G \frac{\epsilon M}{r^2}(1+\frac{p}{\epsilon})(1+\frac{4 \pi r^3 p}{M}){(1-\frac{2GM}{r})}^{-1}- \frac{2\sigma}{r},
\label{TOV}
\end{equation}
while the mass and particles number profiles can be determined from
\begin{eqnarray}
\frac{dM}{dr} &=& 4 \pi \epsilon r^2\nonumber\\
\frac{dA}{dr} &=& 4 \pi \rho r^2{(1-\frac{2GM}{r})}^{-1/2}.
\label{TOV1}
\end{eqnarray}
Here the anisotropic pressure $\sigma$= $p$-$q$. It is obvious that if we set $p$=$q$ ($\sigma$=0 ) in Eq.~(\ref{TOV}), we obtain the standard TOV equation for isotropic star. It is  obvious that the last term in  Eq.~(\ref{TOV}); $\sigma$, represents a kind of force which generated by anisotropy. This force can be directed outward or inward depending on the sign of $\sigma$. Therefore, we can have more massive configuration if $\sigma$ is negative and less massive one if $\sigma$ is positive. The strength and the distribution of the force depend on the magnitude of  $\sigma$ and its profile. Based on this   mechanism, we can support larger masses and radii of neutron star even the EOS of matter is relative soft by adjusting $\sigma$. From above order of magnitude estimations, it can be seen also that electric field effect yields negligible effect on the magnitude of  $\sigma$. The magnetic field  may provide  approximately $\sigma$   $\lesssim $ - 2  $10^{2}$ $\rm MeV fm^{-3}$. However, such a configuration of fluid and magnetic field that close to the  upper bound of this estimation might lead to instability, and stable configurations likely have smaller $\sigma$. While if we used 2 fluids approach, $\sigma$ depends on the value and sign of $b$, a parameter which shows the difference between the four-velocities of both fluids.   

In principle, a realistic $\sigma$ in Eq.~(\ref{TOV}) should be determined from the unified microscopic theory of matter. Unfortunately, as mentioned previously, the appearance of $\sigma$ generate by interplay of many possible microscopic basis in such non trivial way. Then technically it is difficult to derive $\sigma$ from one unified microscopic theory where it can capture effectively all source of anisotropy of matter. Furthermore, mostly the available microscopic models are developed in flat space-time because quantization process of many particles system in curve space-time is known very difficult. While the conventional transferred form of the stress-energy tensor to that for curve space-time, physically may not be too satisfactory (see discussions in Refs. 1-2, and the references therein). The following is a simple example to illustrate the later. Let us consider a star-like object that composing by scalar mesons. They interact each other by exchanging vector mesons~\cite{ASM2009}. If we consider both fields as classical fields, in flat space-time the field of both mesons are spatially homogeneous so that the pressure in  stress-energy tensor in flat space-time is isotropic and the  stress-energy tensor of the matter can be consider as 1 ideal fluid and as the consequence, in conventional view in  curve space-time,  stress-energy tensor of the system  is also 1 ideal fluid and the structure can be obtained by solved  the standard TOV equation for isotropic star~\cite{ASM2009}. However, in fact the spatially homogeneity of mesons fields in flat space-time does not always retain if it is transferred to curve space-time. This can be seen obviously, if we directly calculate the  stress-energy tensor of the matter in curve space-time (see example Ref. 80 for a particular case i.e., with the mass of vector meson is taken to be zero), we can obtain:
\begin{eqnarray}
T^0_0&=& -e^{-2 \nu}[{(\omega_\Phi + g_{V \phi} A)}^2 \Phi^2 +\frac{1}{2} m_V^2 A^2
     + \frac{1}{2}  e^{-2 \lambda} {(\frac{d A}{dr})}^2]\nonumber\\ &-&[m_\Phi^2 \Phi^2 +  e^{-2 \lambda}{(\frac{d\Phi}{dr})}^2]\nonumber\\
T^r_r&=& e^{-2 \nu}[{(\omega_\Phi + g_{V \phi} A)}^2 \Phi^2 +\frac{1}{2} m_V^2 A^2
     - \frac{1}{2}  e^{-2 \lambda} {(\frac{dA}{dr})}^2]\nonumber\\ &-&[m_\Phi^2 \Phi^2 -  e^{-2 \lambda}{(\frac{d\Phi}{dr})}^2]\nonumber\\
T^\theta_\theta=T^\phi_\phi &=& e^{-2 \nu}[{(\omega_\Phi + g_{V \phi} A)}^2 \Phi^2 +\frac{1}{2} m_V^2 A^2
     + \frac{1}{2}  e^{-2 \lambda} {(\frac{dA}{dr})}^2]\nonumber\\ &-&[m_\Phi^2 \Phi^2 +  e^{-2 \lambda}{(\frac{d\Phi}{dr})}^2],
\label{BS}
\end{eqnarray}
where the vector and the scalar fields become inhomogeneous and obey following equations
\begin{eqnarray}
\frac{d^2A}{dr^2}&+&[\frac{2}{r}-(\lambda'+\nu')]\frac{dA}{dr} - e^{2 \lambda} [m_V^2 \nonumber\\ &+& 2  g_{V \phi} \Phi^2] A-  2  g_{V \phi} \omega_\Phi \Phi^2  e^{2 \lambda}=0\nonumber\\
\frac{d^2\Phi}{dr^2}&+&[\nu'-\lambda'+\frac{2}{r}]\frac{d\Phi}{dr}  \nonumber\\ &+&e^{2 \lambda} [ e^{-2 \nu} ({(\omega_\Phi + g_{V \phi} A)}^2-m_\Phi^2]\Phi=0.
\label{BS2}
\end{eqnarray}

 It is obvious from Eq.~(\ref{BS}) that the radial pressure $p ~(T^r_r) $ and the tangential pressure $q ~(T^\theta_\theta)$ of this system are not the same anymore due to the presence of nonzero value of $\frac{dA}{dr}$ and $\frac{d\Phi}{dr} $. How large such effect, is only be known by solving Eqs.~(\ref{BS}-\ref{BS2}) explicitly. But this is already the outside of the scope of this work. Second example of this situation can be seen if we consider a star matter that composed by ideal relativistic free Fermi gas. It is well known, if we quantized in  flat space-time, and transferred its energy-momentum tensor to curve space-time by conventional manner, then we obtain the pressures of this matter is isotropic. However, the author of Ref. 81 demonstrated that if we directly solve the Dirac equation of N free Fermion system in curve space-time, then it is obtained that the pressures of matter becomes anisotropic.

Therefore, based on these twofold difficulties, the expression of $q$ should be modeled. The following conditions are general physical requirements for physically meaningful anisotropic fluid spheres~\cite{HS1997,MH2002}
\begin{enumerate}
\item the energy $\epsilon$ and $p$ should be positive inside the star;
\item the gradients $\frac{d\epsilon}{d r}$, $\frac{dp}{dr}$, and $\frac{dq}{dr}$ should be negative;
\item inside the static configuration the speed of sound should be less than the speed of light, i.e., $ 0 \le \frac{dp}{d\epsilon} \le 1$ and $ 0 \le \frac{dq}{d\epsilon} \le 1$;
\item energy momentum tensor has to be obey the conditions $\epsilon \ge p+ 2q$ and $\epsilon +p+ 2q  \ge 0$;
\item the interior metric should be joined continuously with exterior Schwarzchild metric;
\item  the $p$ must vanish but $q$ may not vanish at boundary r=R of the sphere, but both should be equal at the center of the matter configuration.
\end{enumerate}
Therefore, in  the modeling of $q$ must not deviate from these general requirements.  In this work, we do not propose a new model but only select two  models of $q$  which are quite often studied in communities~\cite{HB2013,DY2012,HIM2011,CHEW1981,HB22013,DHIKV2006,CFV2005}. The first model is obtained by heuristic procedure which allows one to obtain solutions for anisotropic matter from known solutions for isotropic matter~\cite{HB2013,CHEW1981}. See more detail about the procedure used to get this model (HB) of $q$ in Ref. 46. In HB model, $q$ depends on $p$ and $\frac{dp}{dr}$. The second model (DY) is taken from  Refs. 2,45 where $q$ is quasi-local EOS. In DY model, $q$  depends directly on $p$ and local compactness $\frac{2 G M}{r}$ as a quasi-local variable.  See more detail about how to construct quasi-local EOS,  $q$ and its thermodynamics in Refs. 45,84-85. 
The explicit $q$ form of both models, respectively are
\begin{eqnarray}
q &=& p [ 1- \Lambda (\frac{2 M G}{r })],\\
q &=& p +\frac{r}{2}\frac{(1-h)}{h}(\frac{dp}{dr}), 
\end{eqnarray}
where $\Lambda$ and $h$ are anisotropic parameters. For isotropic stars, parameters $\Lambda$ and $h$ should be equal to 0 and 1, respectively. Thus,   the deviation from its reference value  of $\Lambda$=0 or $h$=1 shows the degree of an anisotropy. We need also to note that Refs. 1-2 used polytopes EOS to describe the matter. They also  investigate the effects of the variation of both parameters in the range  of $-2 \le \Lambda \le 2$~\cite{DY2012} and $0.5 \le h \le 1.5$ ~\cite{HB2013} using this EOS. The authors of Ref. 1 found that we can have more massive star if we use $\Lambda <$ 0 while the authors of Ref.~\cite{HB2013} found that more massive star can be obtained if we use positive $h$ and $h <$ 1.  In addition, in the case of DY model, for large magnitude of $\Lambda$ and large masses, NS solution exists for which the energy density $\epsilon$ is not a monotonic function of the radial coordinate but is maximum. While these solutions are dynamically stable~\cite{DY2012,HIM2011},  Ref. 1 shows for the HB model that the  stability of this model increases with the decrease of $h$. 

Here gravitation mass is defined as $M_G$=$M(R)$ and  $A(R)$ is multiplied by atomic unit. 931.50 MeV defines the baryonic mass $M_B$ where  $R$ is the star radius.
Note, general relativity predicts a redshift for photons leaving the surface of the star with strong gravitational field. The gravitational redshift of a non-rotating NS is indicated by 
\begin{eqnarray}
Z= {(1-\frac{2GM}{r})}^{-1/2}-1.
\label{Redsft} 
\end{eqnarray}
By solving Eq.~(\ref{TOV}), numerically, we can study the effects of anisotropic pressure on the  NS gravitational mass-radius relation, NS minimum mass and $Z$ through Eq.~(\ref{Redsft}).

\section{Results and Discussion}
\label{sec_RD}
In this section, we discuss the effects of anisotropic pressure on some
NS properties using one of the parameters in the ERMF model. We adopt the
BSP parameter set because this parameter set provides good descriptions
of the global properties of finite nuclei, and its prediction of nuclear
matter properties  is quite compatible to the prediction from the
heavy ion data. For meson-hyperon couplings, we use conventional SU(6)
symmetry and the experimental values of nuclear matter potential depths
 hyperons in the nuclear matter at $\rho_0$~\cite{JSB_AG}.

\begin{figure}
\centerline{\psfig{figure=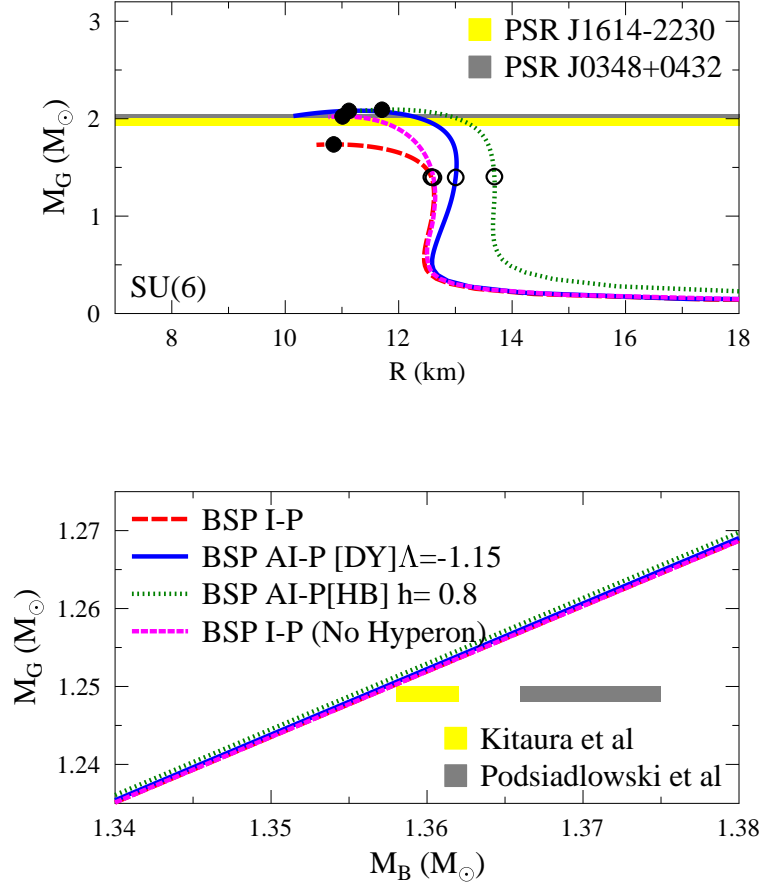, width=12.5cm }}
\caption{Upper panel: The plots for NS gravitational mass-radius relation
for the case of isotropic pressure (I-P) with and without hyperons as well
as the cases of anisotropic pressure (AI-P). The mass of PSR J0348+0432
is taken from Ref.~\cite{Antoniadis13} while the mass of PSR J1614-2230 is from
Ref.~\cite {Demorest10}. Lower panel: The plots for the corresponding
gravitational mass versus baryonic mass for small NS. The shaded boxes are
simulation results by Kitaura {\it et.~al}~\cite{Kita} and Podsidlowski
{\it et.~al} ~\cite{Pot}. DY indicates the result obtained by using
$\sigma$ of Ref.~\cite{DY2012}, and HB is obtained by using $\sigma$ of
Ref.~\cite{HB2013}. The dot and circle markers indicate NS $M_{\rm max}$
and  canonical radius ($M_B$=1.4 $M_\odot$) respectively. Note that for
all NS with hyperons inside its core, the hyperon-meson coupling strength
is determined by using SU(6) symmetry. } 
\label{Fig:MR} \end{figure}

In the upper panel of Fig.~\ref{Fig:MR}, we present the NS gravitational
mass-radius relation for the case of isotropic pressure (I-P) with and
without hyperons (no hyperon) as well as the cases of anisotropic pressure
(AI-P) using DY and HB models, respectively. The mass of PSR J0348+0432
is taken from Ref. 5 while PSR J1614-2230 is from Ref. 4. The dot and
circle markers indicate respectively NS $M_{\rm max}$  and  canonical
radius (radius where $M_G$=1.4 $M_\odot$).  It can be observed that if we
assume the NS matter pressure is isotropic (I-P case), the maximum mass
decreases from $M_G$=2.02 $M_\odot$ to $M_G$=1.74 $M_\odot$
if hyperons are allowed to appear in the NS core.  The value $M_G$=1.74
$M_\odot$ is obviously outside the mass range of PSR J0348+0432 and
PSR J1614-2230. However, if we 
allow that the pressure of NS matter to be anisotropic (AI-P case)
we can obtain maximum mass greater than the masses
of PSR J0348+0432 and  PSR J1614-2230 for $\Lambda \le -1.15$ ($M_G$=2.08
$M_\odot$) for  DY~\cite{DY2012} and $h \le 0.8$ ($M_G$=2.09 $M_\odot$)  for 
the HB~\cite{HB2013} models used. These results can be achieved without
adjusting the hyperons coupling constant, introducing hypothetical
particle like WILB or modifying the nonlinear terms in the strange meson
sector~\cite{BHZBM2012,SA2012,JLC12,WCS2012,SMK2011}. Furthermore, by
assuming that NS matter may have anisotropic pressure, the NS maximum
mass limit higher than 2.1 $M_\odot$ cannot rule out the presence of
exotica in the form of hyperons, boson condensations or quark matter
inside the NS core. For I-P case, the canonical radius with hyperons
is $R_{1.4}$=12.57 km, and $R_{1.4}$=12.61 km without hyperons. These
values are almost the same as there are only very small amount of hyperons
already present in isotropic NS with M$_G$ = 1.4$M_\odot$. For AI-P case,
the $R_{1.4}$ is relatively higher than that in the I-P cases. DY model
predicts $R_{1.4}$=13.00 km for  $\Lambda = -1.15$ while HB model predicts
$R_{1.4}$=13.69 km for $h = 0.8$.  If we compare our  result with the radius
constraint from X-ray bursts, it can be seen that the radii  predicted
by all models considered here are greater than the radius constraint as
obtained by Steiner {\it et. al}~\cite{Stein2010} but smaller than that
of  Sulaimanov {\it et. al}~\cite{Sule2011}. If we compare our result  with
the radius constraint from quiescent low mass X-ray binaries, the radii
result predicted by all models are greater than the radii constraint as
in Guillot {\it et. al}~\cite{Gui2013} but are barely compatible to the
result obtained by Lattimer and Steiner~\cite{LS2013}. However, these
results are compatible with the lower limit of radius of pulsar J0437-4715
i.e., $R$ $>$ 11.1 km within 3$\sigma$ error~\cite{Bog2013}. Therefore,
more accurate radius measurement from other possible sources is needed
to test the reliability of the anisotropic models.

In the lower panel in Fig.~\ref{Fig:MR}, we present gravitational
mass versus baryonic mass for NS with $M_G$ $<$ 1.27  $M_\odot$. It is
obvious in this region that hyperons yield negligible effect while the
effect of anisotropic pressure is apparently insignificant. We note,
the double pulsar J0737-3039 and its interpretation poses a constraint
for this low mass region~\cite{Pot}. The gravitational mass of pulsar B
is measured very precisely while the baryonic mass depends on the mode
of its creation, which can be modeled. The shaded boxes are simulation
results in the form of gravitational mass as a function of baryonic
mass obtained by Kitaura {\it et. al}~\cite{Kita} and Podsidlowski
{\it et. al}~\cite{Pot}. It can be observed that our results by using
BSP parameter set are barely compatible to the result of simulation by
Kitaura {\it et. al.}. We also note that our result is also relatively
close to the calculation result by using the quark-meson coupling model
as obtained by Whittbury {\it et. al}.~\cite{Whitten2013}.

\begin{figure}
\centerline{\psfig{figure=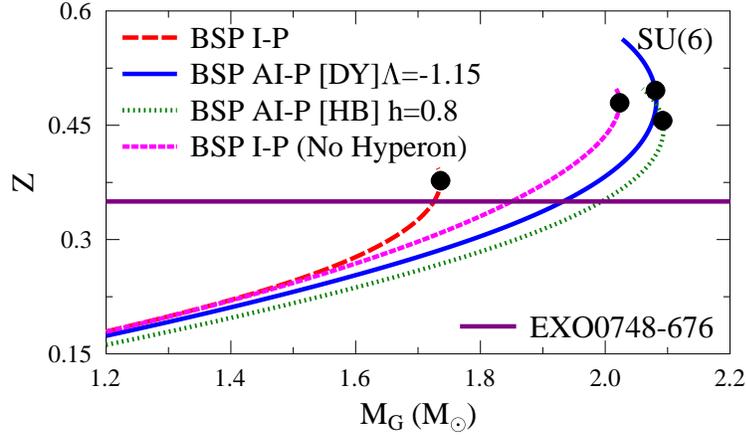, width=12.5cm }}
\caption{The gravitational redshift $Z$ of NS as a function of NS
gravitational mass for the cases of isotropic pressure (I-P) with
and without hyperons as well as the cases of anisotropic pressure
(AI-P). The horizontal line is $Z$ for low mass X-ray binary
EXO0748-676 NS (Ref.~\cite{CPM2002}). The dot marker indicates $Z$ at
$M_{\rm max}$. Note that for all NS with hyperons inside its core, the
hyperon-meson coupling strength is determined by using SU(6) symmetry. }
\label{Fig:RS} \end{figure}

In Fig.~\ref{Fig:RS}, we present the redshift $Z$ as the function
of NS gravitational mass for I-P and AI-P cases by using the HB and
DY models for the EOS of NS core with and without hyperons. The
result is also compared to the observational constraint from
EXO0748-676~\cite{CPM2002}. This constraint implies that the acceptable
EOS should have maximum $Z$ above 0.33~\cite{Lackey2006}. It can be seen
that for all cases, the results are consistent with $Z$ =0.35 for the
mass greater than 1.7 $M_\odot$. These results are quite consistent with
the expected higher masses of accerting stars in X-ray binaries. It can
also be observed that the value of $Z$ at $M_{\rm max}$ in AI-P cases
is higher than that in I-P cases with hyperons. However, this result is
quite similar to the one in I-P cases without hyperons. This is due to the stiffening of EOS 
for the AI-P case. 
The fact that
the radius of NS at $M_{\rm max}$ predicted by  HB with $h = 0.8$ ($R$[$M_{\rm
max}$]=11.70 km)  is slightly greater than that predicted by DY with $\Lambda
= -1.15$ ($R$[$M_{\rm max}$]=11.12 km) is the reason why the  $Z$ at
$M_{\rm max}$ predicted by DY model is higher than the Z in the HB model.

\begin{figure}
\centerline{\psfig{figure=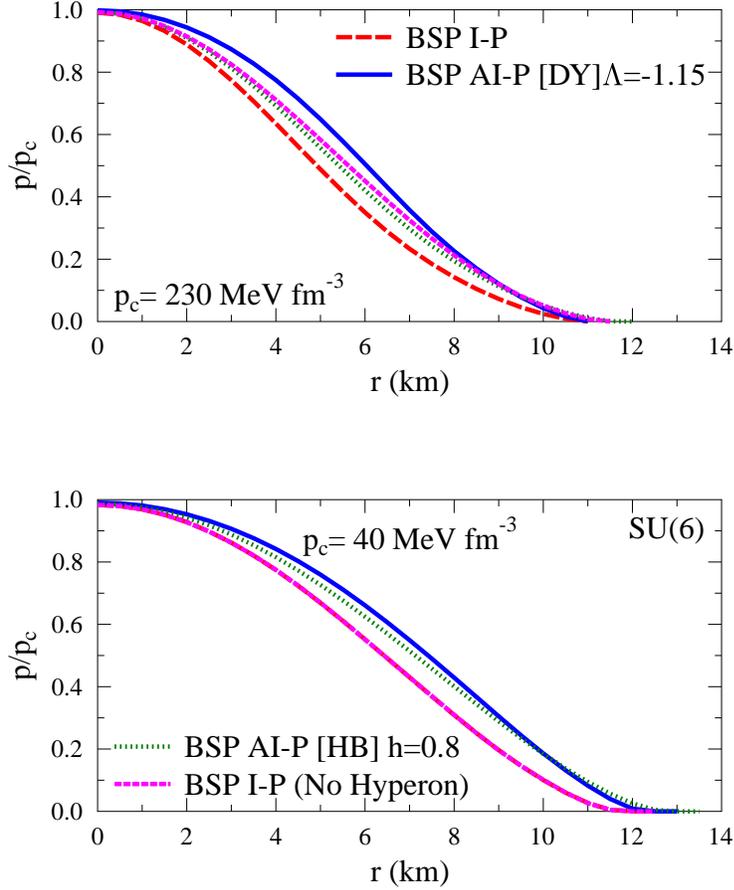, width=12.5cm} }
\caption{The ratio of the radial pressure $p$ to its value at the center
$p_c$ as a function of radial coordinate $r$ for an NS with $p_c$ = 40
MeV fm$^{-3}$ (lower panel) and for an NS  with $p_c$ = 230 MeV fm$^{-3}$
(upper panel) in  the case of isotropic pressure (I-P) with and without
hyperons as well as in the case of anisotropic pressure (AI-P).  Note that
for all NS with hyperons inside its core, the hyperon-meson coupling
strength is determined by using SU(6) symmetry. } \label{Fig:Press}
\end{figure}

\begin{figure}
\centerline{\psfig{figure=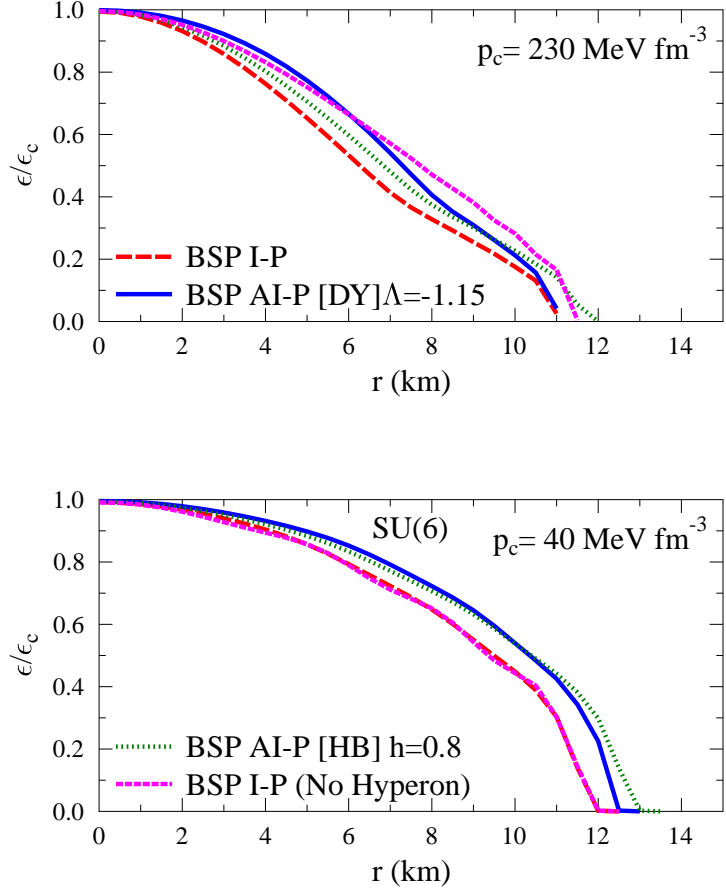, width=12.5cm }}
\caption{Similar to Fig.~\ref{Fig:Press} but for the ratio of energy
density to the value in center as a function of radial coordinate $r$
for an NS with $p_c$ = 40 MeV fm$^{-3}$ (lower panel) and for an NS with
$p_c$ = 230 MeV fm$^{-3}$ (upper panel).} 
\label{Fig:Eden} \end{figure}

To investigate further, in Figs.~\ref{Fig:Press} and~\ref{Fig:Eden}
we plot
the ratio of the radial pressure $p$ to its value at the center $p_c$
and the ratio of the energy density $\epsilon$ to its value at the
center $\epsilon_c$ as a function of radial coordinate r. 
The results plotted in the lower and  upper panels correspond to the
central pressure $p_c$ = 40 and 230 MeV fm$^{-3}$, respectively. For $p_c$ = 40 MeV
fm$^{-3}$, BSP AI-P [DY] has $M_G$ = 1.47 $M_\odot$,  BSP AI-P [HB] has
$M_G$ = 1.55 $M_\odot$, BSP I-P with and without hyperons yield identical
mass  i.e., $M_G$ = 1.19 $M_\odot$, because  for these cases hyperons
have not yet appeared in NS and the effects on the pressure profile are
only due to the anisotropy.  For  $p_c$ = 230 MeV fm$^{-3}$, BSP AI-P
[DY] yields $M_G$ = 2.1 $M_\odot$,  BSP AI-P [HB] yields $M_G$ = 2.1
$M_\odot$, and   BSP I-P yields    $M_G$ = 1.7 $M_\odot$ but BSP I-P
(no-Hyperons) has $M_G$ = 1.9 $M_\odot$. In this case, we can observe
the effects of the interplay of hyperons and anisotropy roles.

\begin{figure}
\centerline{\psfig{figure=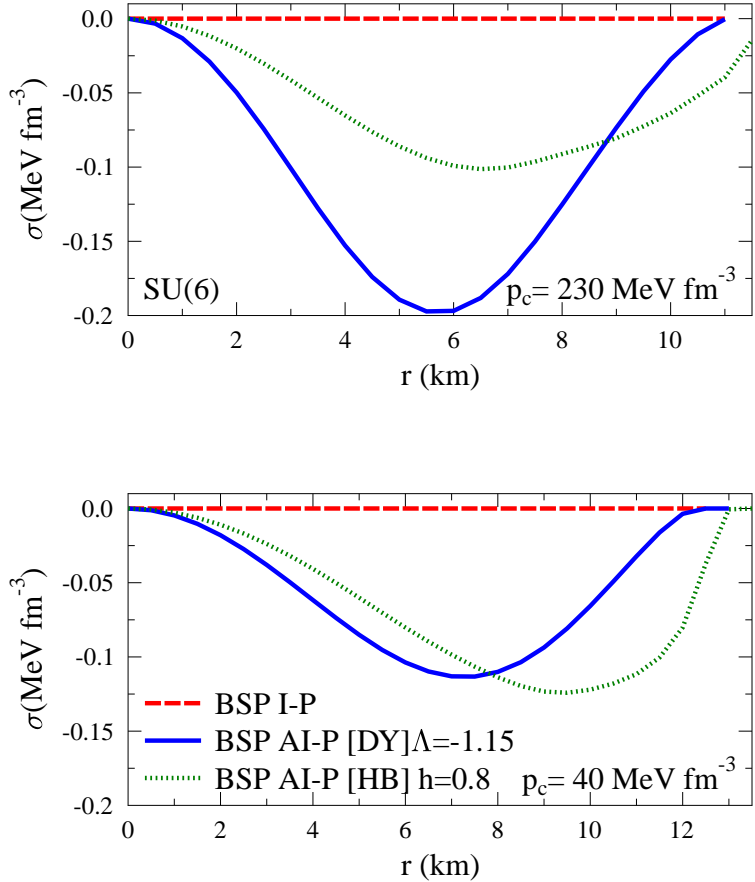, width=12.5cm }}
\caption{Similar to Fig.~\ref{Fig:Press} but for the anisotropic
pressure $\sigma$ of NS with hyperons inside its core as a function of
radial coordinate $r$  for an NS with $p_c$ = 40 MeV fm$^{-3}$ (lower
panel) and for an NS with $p_c$ = 230 MeV fm$^{-3}$ (upper panel). }
\label{Fig:SR} \end{figure}

\begin{figure}
\centerline{\psfig{figure=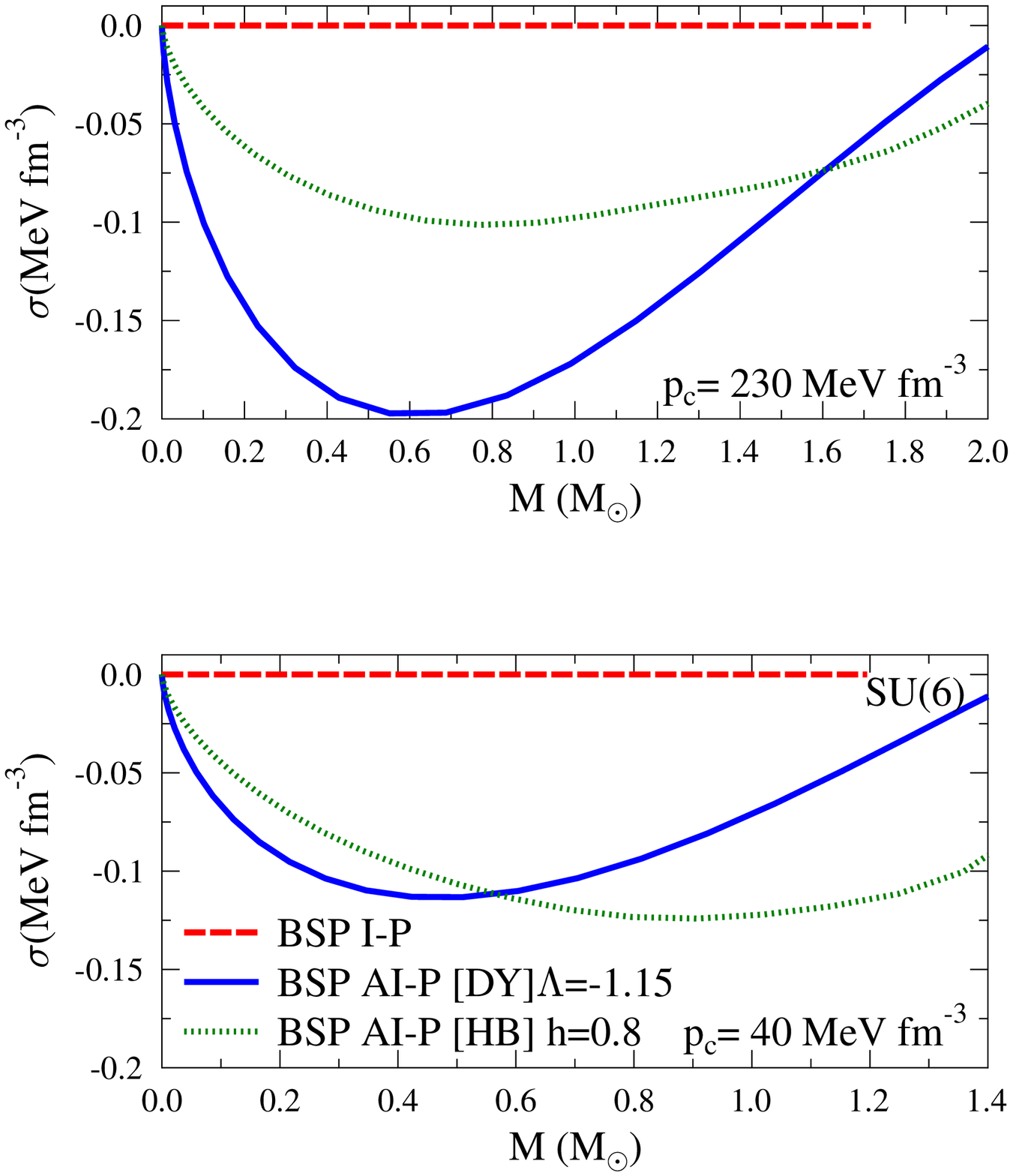, width=12.5cm }}
\caption{Similar to Fig.~\ref{Fig:Press} but for the anisotropic
pressure $\sigma$ of NS with hyperons inside its core as a function
of its mass distribution $M(r)$ for an NS with $p_c$ = 40 MeV fm$^{-3}$
(lower panel) and for an NS with $p_c$ = 230 MeV fm$^{-3}$ (upper panel).}
\label{Fig:SM} \end{figure}

The absence of hyperons slowly decreases $p$
and $\epsilon$ with increase in the distance from the center of a NS. The
role of anisotropic pressure is the same as that of the case with no  hyperons,
i.e., $p$ and $\epsilon$ decreases slowly with increase in the distance
from the center of a NS. But at low $p_c$ or $\epsilon_c$ values, the
anisotropic pressure still yields pronounced impact. However,
effect of  AI-P in slowing down the rate of decreasing $p$ and $\epsilon$
due to application of different anisotropy models appears more significant
at relatively high radial $p_c$ or $\epsilon_c$ values. The latter can be
understood by observing the plots of anisotropic pressure $\sigma$ of NS
with hyperons from the HB and DY models as a function of radial coordinate
$r$ and mass distribution $M$ which are shown in Figs.~\ref{Fig:SR} and
~\ref{Fig:SM}. It can be observed that unlike the HB model with almost
constant   absolute value of maximum anisotropic pressure $\sigma$,
the prediction using the DY model is rather sensitive to the value of
$p_c$. DY model predicts greater $\sigma$ for higher $p_c$. For $p_c$ = 40
MeV fm$^{-3}$, the DY model predicts  absolute value of maximum $\sigma$
$\sim$ 0.1 MeV fm$^{-3}$ at $M \sim$ 0.4  $M_\odot$ and $r~\sim$ 7 km
while the HB model predicts the same absolute value of maximum  $\sigma$
at  $M~\sim$ 0.8  $M_\odot$ and $r$ $\sim$ 10 km. For $p_c$ = 230 MeV
fm$^{-3}$ on the other hand, the DY model predicts absolute value of
maximum  $\sigma$ $\sim$ 0.2 MeV fm$^{-3}$ at  $M~\sim$ 0.6  $M_\odot$
and $r \sim$ 6 km while the BD model predicts the absolute value of
maximum  $\sigma$ $\sim$ 0.1 MeV fm$^{-3}$ at $M~\sim$ 0.7  $M_\odot$
and  $r~\sim$ 6 km. To this end, it is clear that if we use the EOS of
NS core without hyperons as a reference, the presence of hyperons in
NS core suppresses the EOS stiffness by decreasing $p$ or $\epsilon$
but by allowing the NS pressure anisotropic, the stiffness of the EOS is
pulled back to relatively higher $p$ or $\epsilon$ so that the $M_{\rm
max}~\ge$ 2.1  $M_\odot$ can be reached. However, some details such
as the radius prediction still depend significantly on the  model used
to describe the anisotropic pressure. Therefore, beside the existence of
more accurate NS observable data like mass, radius etc, it seems that
systematic reassessment of the consistency of the anisotropic pressure
models offer in the literature with all possible physical requirements
for anisotropic star including the star stability using ``realistic
matter EOS'' are important to select the most physically acceptable
model. However, this is already outside the scope of this work. We leave
it as our next project.

\section{Conclusion}
\label{conclu}
We have studied the effect of anisotropic pressure on the gravitational
mass, baryonic mass, radius and the redshift of static NS with the
presence of hyperons in NS core. The BSP parameter set of the ERMF
model~\cite{ASR2012} and standard SU(6) symmetry as well as the
experimental value of nuclear matter potential depths  are used to
determine the hyperon-meson coupling constants~\cite{JSB_AG}. To describe
the anisotropic pressure we adopt two known models in the literature,
namely DY~\cite{DY2012} and HB~\cite{HB2013}  models. The effect
of anisotropic pressure on NS matter is mainly to increased stiffness
of the NS EOS. This effect can compensate the softening of the NS
EOS due to the presence of hyperons. Without further adjusting the
hyperons coupling constant, introducing hypothetical particle
like WILB or modifying the nonlinear terms in strange meson
sector~\cite{BHZBM2012,SA2012,JLC12,WCS2012,SMK2011}, we can easily
obtain larger $M_G$ than the masses of PSR J0348+0432 and  PSR
J1614-2230 if $\Lambda \le -1.15$ for DY~\cite{DY2012} and $h \le 0.8$
for HB~\cite{HB2013} models are used. In anisotropic NS, the maximum mass
limit higher than 2.1 $M_\odot$ cannot rule out the presence of exotica
in the form of hyperons, boson condensations or quark matter inside
the NS core. We have found relatively large canonical radius. DY model
predicts $R_{1.4}$=13.00 km for  $\Lambda = -1.15$ and HB model predicts
$R_{1.4}$=13.69 km for $h = 0.8$. We also found that the relation between
gravitation mass and  baryon NS mass remain practically unaffected for
$M_G$ $<$ 1.27  $M_\odot$ due to the anisotropic pressure. Furthermore,
the anisotropic pressure can increase the value of  $Z$ at maximum
mass. We also show that the minimum mass, radius and the redshift
predictions of anisotropic NS are quite compatible with the recent
observational constraints.

\section*{Acknowledgments}
A. S. acknowledges the support given by Universitas Indonesia through Hibah Klaster Riset 2014 (No:1709/H2.R12/HKP.05.00/2014). A. S. also want to express many thanks to Prof. B. K. Agrawal for checking thoroughly the English of this paper.

\end{document}